\documentclass[conference]{IEEEtran}
\ifCLASSINFOpdf
\else
   \usepackage[dvips]{graphicx}
   \DeclareGraphicsExtensions{.eps}
\fi
\begin{document}
\title{Magnetohydrodynamics on Heterogeneous architectures: a performance comparison}

\author{\IEEEauthorblockN{Bijia Pang}
\IEEEauthorblockA{Department of Physics\\
University of Toronto\\
Toronto ON, M5S 1A7, Canada\\
Email: bpang@physics.utoronto.ca}
\and
\IEEEauthorblockN{Ue-li Pen}
\IEEEauthorblockA{Canadian Institute for Theoretical Astrophysics\\
University of Toronto\\
Toronto ON, M5S 3H8, Canada\\
Email: pen@cita.utoronto.ca}
\and
\IEEEauthorblockN{Michael Perrone}
\IEEEauthorblockA{IBM TJ Watson Research Center\\
Yorktown Heights, NY 10598, USA\\
Email: mpp@us.ibm.com}}


%Telephone: (800) 555--1212, Fax: (888) 555--1212}


\maketitle

\begin{abstract}
%\boldmath
We present magneto-hydrodynamic simulation results for heterogeneous systems. 
Heterogeneous architectures combine high floating point performance many-core units hosted in conventional server nodes.  Examples include Graphics Processing Units (GPU's) and Cell. 
They have potentially large gains in performance, at modest power and monetary cost.

We implemented a magneto-hydrodynamic (MHD) simulation code on a variety of  heterogeneous and multi-core  architectures --- multi-core x86, Cell, Nvidia and ATI GPU --- in different languages, FORTRAN, C, Cell, CUDA and OpenCL.   
We present initial performance results for these systems.  
To our knowledge, this is the widest comparison of heterogeneous systems for MHD simulations.
We review the different challenges faced in each architecture, and potential bottlenecks.  
We conclude that substantial gains in performance over traditional systems are possible, and in particular that is possible to extract a greater percentage of peak theoretical performance from some systems when compared to x86 architectures.
\end{abstract}


% \fi
\IEEEpeerreviewmaketitle

\section{Introduction}

Magneto-hydrodynamics (MHD) studies the dynamics of magnetized conducting fluids.
If there is no magnetic field present, the problem reduces to traditional
fluid dynamics.   However, in most astrophysical settings for instance, the fluids are highly
conductive, and observed to be magnetized. Electro-motive forces
generated by magnetic fields will modify the flow, which will in turn
affect the field.  As a result, one has to solve both the
Euler equations and Maxwell's equations simultaneously.

The MHD equations are nonlinear, and cannot in general solved analytically.
Thanks to the increasing power of computers, three dimensional
simulations can be used to model these dynamics numerically.
Numerical simulations are crucial both for understanding the theory of
such fluids, and for use in directing real world experiments.  However, in order to
achieve realistic parameter regimes to solve real world problems,
numerical experiments must push the limits of computational hardware
resources.  

Increasingly, considerations of compute power per watt or per dollar mean that
new architectures are being considered to perform these calculations.   In particular,
we examine here heterogeneous systems, which consist of  
two different kinds of processors; one or more general-purpose
conventional processors which control the overall computation,
and specialized, usually multi-core, processor units to which the numerically intensive computing
is offloaded  \cite{shan2006heterogeneous}.  There are several heterogeneous
platforms in use currently and we will focus on Cell/B.E. \cite{ibmcell:website} and
graphics processing units (GPU) in this paper, and use a multi-core x86 system
for comparison.  

Complicating comparison between these systems is that typically they are programmed
using different platform-specific languages.    The Open Computing Language (OpenCL) \cite{OpenCL:website}, is cross-platform application programming interface (API), which is
designed for heterogeneous systems, including GPU and Cell, has been released by Khronos Group,
and which ameliorates this problem to a degree; however we will see here that at the current time this does not solve the problem, and we must in general still use the platform-specific languages for performance. 

We discuss our reference implementation of a solver for the MHD equations in \S II; in \S III we discuss our implementation on several architectures.    We summarize our results in \S IV and discuss future work; in \S V we conclude.

\section{MHD equations and the algorithms}

\subsection{MHD equations}

As with hydrodynamics, the MHD equations conserve mass, momentum, and energy; in addition,
there is an induction equation in which field motions can `stretch' field lines, which by `magnetic flux freezing' are frozen in to fluid elements\cite{nla.cat-vn2032203}:

\begin{equation} \label{MHD_1}
\partial_{t}{\rho}+{\nabla}{(\rho \vec{v})}=0 
\end{equation}
\begin{equation} \label{MHD_2}
\partial_{t}{(\rho\vec{v})}+{\nabla}(\rho\vec{v}\vec{v}+P_{\ast}{\delta}-\vec{b}\vec{b})=0
\end{equation}
\begin{equation} \label{MHD_3}
\partial_{t}{e}+\nabla[(e+P_{\ast})\vec{v}-\vec{b}\vec{b}\cdot\vec{v})]=0
\end{equation}
\begin{equation} \label{MHD_4}
\partial_{t}\vec{b}=\nabla\times(\vec{v}\times\vec{b})
\end{equation}
\begin{equation} \label{MHD_5}
\nabla\cdot\vec{b}=0
\end{equation}

Here for numerical convenience the magnetic field $b$ is normalized by a factor of $\sqrt{4\pi}$.
$P_{\ast}$ is total pressure, which equals to the sum of the gas pressure $p$ and the magnetic pressure $b^2/2$; $\rho$ and $e$ are the mass and total energy densities,
where the latter is the sum of kinetic energy ($\rho v^{2}/2 $), 
internal energy ($p/(\gamma-1)$),
and magnetic energy ($b^2/2$).

\subsection{Algorithm}

There are many algorithms for solving these equations, which we will not attempt
to review here.  We follow the approach of \cite{2003ApJS..149..447P} in this paper, as the conciseness of its implementation lends itself to re-implementation for the different architectures, and its memory-access patterns are an excellent match to the heterogeneous architectures discussed here.

The method is a second-order accurate (in space and time) high-resolution
total variation diminishing (TVD) \cite{harten1997high} scheme.
The kinetic, thermal, and magnetic energy are conserved identically and
there is no explicit magnetic or viscous dissipation.
The TVD constraints result in non-linear viscosity and resistivity
on the grid scale.  The TVD constraint allows the capture of shocks
for compressible flows, where the flow becomes discontinuous.

The code solves the magnetic component and fluid dynamics separately.  The former is solved by a two-dimensional advection-constraint step \cite{2003ApJS..149..447P},  while for the latter, a monotone upwind scheme for conservation laws (MUSCL) is used for a one dimensional fluid advection update \cite{2003PASP..115..303T}. The time step update is based on Courant-Friedrichs-Lewy (CFL) constraint,
which ensures that the fastest wave can't travel for more than one grid space in a single time step.
The approach is `dimensionally split' in the sense that updates are first made along the $x$ direction, then $y$, and then $z$; memory transposes are used to reorient the grid between each sweep.  This
both greatly simplifies the numerical kernel (which only has to be implemented once) and ensures regular memory access for each sweep.

The dimensional splitting reduces the fluid update to one dimensional dynamics:
\begin{equation} \label{1Dadv}
\partial_{t}{\vec{u}}+{\nabla}_{x}{\vec{F}}=0 .
\end{equation}
This is discretized into finite volumes, ensuring conservation.   The fluxes are calculated 
using MUSCL, a first-order upwind scheme with a second-order TVD (Van Leer limiter) correction.
Time integration is performed using a second-order Runge-Kutta scheme.  To solve the complex upwind problem that is involved with momentum and energy fluxes,
relaxing TVD \cite{Jin95therelaxation} is used for the Euler equations.

The magnetic update is reduced to a two dimensional advection-constraint step consistent with Eq.~\ref{MHD_4} and to ensure the constraint given by Eq.~\ref{MHD_5}.   In constrained transport \cite{1988ApJ...332..659E}, one stores the magnetic flux at the cell face, which can then be used to accurately maintain a zero divergence of magnetic field.  

In addition,  \cite{2003ApJS..149..447P} proposed not storing all the computed electromotive forces (EMFs) and just applying the individual pieces of the EMF for advection-constraint steps.  This can save a significant amount of memory, and in addition, reduce unnecessary memory access.  As a result, the code is very memory efficient, and transposing the grid in memory between sweeps ensures short strides along sweep directions and thus low memory-access latency.     One must remain aware of
grid-imposed data dependencies of the method, however.  The one dimensional fluid update stencil is a standard 7-point stencil requiring data from all 4-neighbouring `pencils' in the direction of a sweep; the magnetic update, to ensure the consistency of the magnetic field constraint, in addition needs the adjacent `pencils' to be updated by the flux.   

\section{Implementation on heterogeneous systems}

Heterogeneous systems have processors for different roles.
As a result, a new memory system design is needed,
which is the challenge for the programmers.

In this section, we will discuss our implementation and performance results of the MHD scheme described above on different platforms: multi-core x86, Cell/B.E., a Nvidia GPU and an ATI GPU. Each platform has corresponding languages or libraries: OpenMP for multi-core x86, Cell programming for Cell blade, CUDA for the NVidia GPU, and OpenCL for the ATI GPU.  

In all cases, we implement the full 3D version of the method described above.   
Our performance tests consist of measuring the time taken to evolve a 3D domain of varying size ($16^3$, $32^3$, $64^3$, and $128^3$ zones) by one timestep (only evolution step and no extra memory transfer is included); by varying the size of the domain we can see the effects of overhead such as memory transfer.  Note that all calculations in this paper are performed at single precision to make comparisons more readily meaningful.   All the timing data has units of milliseconds.

\subsection{Multi-core x86}
As a basis of comparison, we first examine the performance of the original FORTRAN code on a multi-core x86 architecture.  We use two Intel Xeon(R) E5506 CPU @ 2.13GHz, each with 4 processor cores,  for this experiment.   

Parallelization is done with OpenMP.   Programming OpenMP is straightforward: the programmers only need to add some lines to the loops and the API will partition the loop automatically.  The
original version of the code under consideration here already had OpenMP parallelization, incurring only a minimal overhead in coding length or complexity.    The parallelization is done over 2D slabs, with parallelization occurring over the outermost loop in the solvers.

\subsubsection{Result}

Data for different box sizes are provided in table \ref{table:x86},  with the numbers inside the brackets indicating the number of cores.   For problem sizes larger than $16^3$, a steady 6.7 times speedup is achieved.

% use the cell cluster from ibm
\begin{table}

\centering
\caption{Performance on the multi-core x86 for different box sizes; timings in milliseconds.
x86(1) refers to single-core performace;
x86(8) to 8.
}
%    \advance\extrarowheight-5pt
\begin{tabular}{|c|c|c|c|c|}
\hline
 & \multicolumn{4}{|c|}{Domain size} \\
\hline
Architecture & $16^3$ & $32^3$ & $64^3$ & $128^3$\\
\hline
x86(1) & 17.8 	& 140 	& 1096 	& 8770 \\
x86(8) & 4.0 	& 20.7 	& 163.6 & 1315 \\
\hline
speedup (8:1) 	& 4.4 	& 6.7  	& 6.7 	& 6.7 \\
\hline
\end{tabular}
\label{table:x86}
\end{table}

\subsection{Cell}
The Cell Broadband Engine (Cell/B.E.) is a collaboration of Sony, Toshiba and IBM.
The original design purpose was for a gaming machine, the Sony's Playstation 3;
however, it is also a good candidate for high performance computing due to its specialized multi-core architecture.
Cell/B.E.'s design, a combination of one Power Processor Element (PPE) and eight Synergistic Processing Elements (SPE),
is to overcome three walls -- the power wall, memory wall and frequency wall \cite{arevalo2008programming}.

The PPE is a 3.2GHz PowerPC-like processor,
and is used to control the eight 3.2GHz SPEs, which are used for data intensive computing.
An SPE can perform four single-precision floating-point operations in a single clock cycle.
With dual pipelines,
this gives $3.2\times4\times2=25.6$~Gflops peak performance for single precision on one SPE \cite{arevalo2008programming}.
There are three levels of memory:
the PPE's main storage,
the SPE's 256kB SRAM local storage, 
and the SPE's 128-bit 128-entry unified register file.
It is the programmers' job to handle the Direct Memory Access (DMA) 
to transfer the data between PPE and SPE.
The transfer is performed on Element Interconnect Bus (EIB),
a high speed internal bus which has 204.8 GB/s peak data bandwidth\cite{chen2007cell}.

Cell processors have a high-level C-like programming language. 

\subsubsection{Parallelization/Partition}

In the first stage in the parallelization, the PPE assigns the threads/memory to the SPEs and performs synchronization.   Once the calculation begins, the PPE will no longer be involved in the calculation,
and all work is done by the SPEs.  DMA is used to transfer data between main memory and local storage.  Since the PPE is not used during the calculation, the signal-notification channel is used for synchronization.  One SPE is assigned as the master. Once a synchronization point is reached,
the slave SPEs send a message to the master SPE.  Upon receiving all the messages, the master initializes slaves using a binary synchronization tree.

The fluid updates are performed along one dimensional pencil of grid points (e.g. X direction), transferred separately to each SPE to calculate; this makes best use of the fairly modest 256kB limit of local storage on each SPE.  By ensuring the domain sides are always multiples of 4,  the starting address of each transfer is correctly aligned.  We follow the update order from the FORTRAN version for the fluid part.  For the magnetic update, any pencil that sits in one SPE has to update the pencil next to it (both Y and Z directions).  As a result, we separate the update of the magnetic part into four sub-functions. The intermediate value to be updated by the next pencil is sent back to the PPE.
After the synchronization of the former function,  the value is sent to the SPEs to finish the update.

Implementing the grid transpose efficiently requires some care.  
For every DMA transfer, the start of the address has to be aligned to 16 bytes.  
To achieve higher performance, the data for each transfer should be approaching 16kB.   
For the regular memory accesses involved in the fluid and magnetic update this is straightforward; but balancing these constraints for the non-continuous memory access of the transpose is more difficult.   
As a result, DMA lists, commands that can cause execution of a list of transfer requests, are used for this task.  For every SPE, there are $16^3$ cube data elements for one component transfer by DMA list. The incoming lists hold the starting address of a two dimensional plane data arrays and the data size. After the transfer inside the SPEs, 
the out-going lists hold the starting address of the after-transpose plane data arrays and the same data size. Because the size is a multiple of 4, with at least single precision (4 bytes), the starting address is always a multiple of 16 bytes.

\subsubsection{Optimization}

Further performance gains can be achieved by taking advantage of SIMD capabilities of the SPEs, and overlapping communication and computation.

To exploit the SIMD capabilities of the SPEs, our code's data structures are arranged as a structure-of-arrays (SOA), which means that the different components of the fluid and magnetic parts are stored in different arrays.  For every SIMD operation in single precision,  one component of the adjacent four cells will be calculated.

To overlap communications and computations, since there is no cache on the SPE and we want to keep the SPEs busy with computing, double buffering is used to hide the memory latency between the PPE and the SPEs.

\subsubsection{Results}
Data for different box sizes are provided in table \ref{table:Cell},
with the number 16 inside the brackets indicating the speed-up ratio for 16 SPEs.

% use the cell cluster from ibm
\begin{table}
\centering
\caption{Cell performance while using PPE or varying numbers of SPEs for different box sizes; timings in milliseconds.
}
%    \advance\extrarowheight-5pt
\begin{tabular}{|c|c|c|c|c|}
\hline
 & \multicolumn{4}{|c|}{Domain size} \\
\hline
Architecture & $16^3$ & $32^3$ & $64^3$ & $128^3$\\
\hline
PPE 		& 52 	& 448 	& 3745 	& 32300\\
1 SPE  		& 22.3 	& 163.8 & 1257 	& 9901 \\
4 SPE  		& 6.5 	& 43.8 	& 327 	& 2607 \\
16 SPE 		& 3.5 	& 14 	& 112 	& 864  \\
\hline
speedup (16 SPE:PPE) 	& 14.9 	& 32.0 	& 33.4 	& 37.4 \\
speedup (16 SPE:1 SPE) 	& 6.4	& 11.7 	& 11.2 	& 11.5 \\
\hline
\end{tabular}
\label{table:Cell}
\end{table}

\subsection{Nvidia GPU}

Graphics Processing Units (GPU) were originally developed for 3D graphics rendering, but their naturally parallel architecture is also suitable for high performance computing.  Current GPUs already use unified shaders for rendering, and these shaders are what we call `cores' or
`stream processors' for GPU computing.

The GTX 260 (192 cores) is used in our tests. There are 8 thread processing clusters (TPC),
which contain 24 streaming multiprocessors (SM),
partitioned into 8 scalar thread processors (TP) running at 1.3GHz (3 single-precision operations per clock cycle). This gives the GTX 260 a peak performance for single precision of:

$8\times24\times1.3\times3=748.8$~Gflops \cite{gtx260:website}.
There are three levels of memory:
1GB of GPU global memory,
16kB of shared memory on each block (i.e. SM),
and 16384 32-bit registers on each block;
In addition, there is read-only memory in the form of constant and texture memory.
The bandwidth between global memory and in-block memory is 141 GB/s,
while the CPU and GPU are connected by PCI-e, which has 8GB/s bandwidth.

For this architecture, we re-implement the MHD solver using the 
Compute Unified Device Architecture (CUDA), a high level C-like language,
which can be used to program on any Nvidia GPU after G80.

\subsubsection{Parallelization/Partition}

For this architecture, it is the CPU which initializes the work, assigns the threads/memory, and performs necessary synchronization.  To minimize the impact of the relatively low bandwidth over PCI-e,
no more data transfer is performed after transferring the initialized data to GPU global memory.
However, there is still the long latency (several hundred cycles) of fetching data from the card's global memory to the arithmetic  units, which must be hidden by oversubscribing the cores.  

CUDA uses SIMT (Single Instruction, Multiple Thread), which means every thread in the same block executes  the same instruction at the same time.   SIMT is different from SIMD in that the width (the number of threads) is not fixed,  which will affect the available number of registers (which is fixed per SM) available per thread.

In our implementation for this architecture, each CUDA block of threads is assigned one one-dimensional pencil, and the corresponding data is copied into the block's shared memory.   Each thread within the block corresponds to one zone.  Synchronization is provided inside a block, and if synchronization among blocks is needed, we return all blocks back to CPU control by ending the CUDA kernel.

To further reduce latency resulting from access to global memory,
we modify the magnetic update by staggering the updates;  first update the odd indices of the blocks,
and subsequently the even indices.  The reading/writing of intermediate flux can be avoided, and about a 10$\%$ speed up is achieved.

Finally, the CUDA SDK provides examples for performing transposes, which are used and modified for our purposes here.   Because the memory transpose is three dimensional, the data size is limited by the shared memory per block. In our simulation,  only $8^3$ grid points of only one component of either fluid and magnetic field are transposed at a time.  We found this to be the best balance between shared memory and data transpose size.

\subsubsection{Optimization}

We can further improve the performance on this architecture by being aware of the underlying memory architecture, and choosing block sizes to maximize occupancy.

Because of the size of the stencil, and the structure of the magnetic field update, adjacent cells are needed for evolving any zone.    Repeated access to global memory is avoided  by using shared memory in CUDA to cache the needed values.   We did not use constant, texture or pinned memory, as there is no large amount of `read-only' data which could benefit from being stored here.

The global memory access by the updates is automatically coalesced by the memory transposes, so needs no special work in this implementation.

A further concern is occupancy -- keeping each SM as fully occupied with thread blocks as possible.
Occupancy is the ratio of active warps to maximum warps in a block.   Increasing occupancy may not lead to good performance directly,  but a low occupancy will certainly not hide memory latency well.
Three factors --- threads per block, shared memory and register usage --- affect the occupancy.  Empirically, we found that organizing the thread blocks by pencils, and assigning between 128 and 192 threads (and thus zones) per block to maximize performance.     Further improvements in occupancy is limited by register number for the fluid evolution and shared memory for the magnetic evolution.

\subsubsection{Result and comparison with previous work}

Timing data for different domain sizes are provided in table \ref{table:Nvidia}.   For sufficiently large domains, we achieve a factor of 100 speedup compared to a single-core x86.  

For this architecture, there is other work that can be used to gauge the efficiency of our implementation.
Two other groups (\cite{wong2009magnetohydrodynamics}, \cite{Schive:2009hw}) have used CUDA to implement  Pen's \cite{2003ApJS..149..447P} TVD code for MHD or pure hydrodynamics.
In \cite{wong2009magnetohydrodynamics}, they used CUDA for MHD and they achieved a speed-up of 84 times in 3D,  on a GTX 295 (480 cores) over an Intel Core i7 965 3.20GHz.   In comparison, our 105 speedup with a less-capable GPU (GTX 260, 192 cores) and on a lower-clock speed CPU (Xeon(R) E5506 2.13GHz) seems at least comparable.

In \cite{Schive:2009hw}, a relaxing TVD scheme was used for three dimensional hydrodynamics.
Furthermore, adaptive mesh refinement (AMR) and a multi-level relaxation scheme were used,
and this was applied to a multi-GPU cluster system.   Since this setup is significantly different from our own, no direct comparison is presented here. They state that their speed-up is 12.19 for 1 GPU.

\begin{table}
\centering
\caption{x86 vs NVidia GPU performance  for different box sizes; timings in milliseconds.
}
%    \advance\extrarowheight-5pt
\begin{tabular}{|c|c|c|c|c|}
\hline
 & \multicolumn{4}{|c|}{Domain size} \\
\hline
Architecture 	& $16^3$ 	& $32^3$ 	& $64^3$ 	& $128^3$\\
\hline
x86(1) 		& 17.8 	& 140 	& 1096 	& 8770 \\
Nvidia (CUDA) 	& 1.36 	& 2.8 	& 11.2 	& 83.0 \\
Nvidia (OpenCL) & 2.4 	& 4.2 	& 15.2 	& 109\\
\hline
Speedup (CUDA:x86)		& 13.1 & 50 & 97.9 & 105.7 \\
\hline
\end{tabular}
\label{table:Nvidia}
\end{table}

\subsection{ATI GPU}
The ATI GPU uses superscalar cores (shader), a modification from SIMD.
One superscalar structure contains one 4D vector and one 1D scalar,
which means in one cycle, it can do one 4D operations and one 1D operation.
To compensate for the insufficient power of scalar computing,
more cores are added onto the chips, e.g. the Radeon HD 5800 has 1600 cores.

We used an ATI HD 5870 for our simulation. 
The 1600 0.85 GHz shader cores are located in 20 SIMD units, and each SIMD unit has 80 cores.
The 4D+1D core can perform two single-precision operations per clock cycle,
which gives the ATI HD 5870 peak performance for single precision as:
$1600\times0.85\times2=2720$~Gflops \cite{ati5870:website}.
There are three levels of memory:
1GB of GPU global memory,
32kB of shared memory per block (i.e. SIMD unit),
and 16384 128-bit registers per block.
The bandwidth between global memory and in block memory is 153.6 GB/s,
while the CPU-GPU's bandwidth is the same as Nvidia.

For this architecture, we re-implement the MHD solver using OpenCL.

\subsubsection{Parallelization/Partition/Optimization}
The parallelization is similar to Nvidia GPU,
except that we vectorize the code to get the maximum performance.
There are both SIMD and SIMT units in ATI GPUs,
and each thread manipulates the data itself, making cross-grid calculation impossible.
As a result, we use structure-of-arrays (SOA), instead of AOS in Cell.
We store the first four of five fluid components as a `float4',
and leave the last one as a `float'.
For the magnetic part, we package the components as a `float4',
leaving the fourth element of magnetic array unused.
For the memory transpose function,
we use two subroutines:
the first for first four components of fluid,
and the second for the fifth component of fluid and magnetic components.
Otherwise, there are few differences for parallelization and optimization between CUDA and OpenCL.

\subsubsection{Results}
Data for different box size are provided in table \ref{table:ATI}.

\begin{table}
\centering
\caption{x86 vs ATI GPU performance  for different box sizes; timings in milliseconds.
}
%    \advance\extrarowheight-5pt
\begin{tabular}{|c|c|c|c|c|}
\hline
 & \multicolumn{4}{|c|}{Domain size} \\
\hline
Architecture & $16^3$ & $32^3$ & $64^3$ & $128^3$\\
\hline
x86(1) 	& 17.8 	& 140 	& 1096 	& 8770 	\\
ATI GPU & 10 	& 26 	& 37 	& 128 	\\
\hline
Speedup (ATI:x86) 	& 1.78 	& 5.4 	& 29.6 	& 68.5 	\\
\hline
\end{tabular}
\label{table:ATI}
\end{table}

\section{Comparative Results and Discussion}

\subsection{Results}
We compare different architecture results by four criteria:

1. Code speed-up: speed up ratio on the heterogeneous architecture compared to a single core x86;

2. Fractional speed-up: ratio of the speed up ratio (heterogeneous to single-core x86) to theoretical peak performance ratio (heterogeneous to single-core x86);

3. Floating-point operations per second (FLOPS) fraction: ratio of actual FLOPS to theoretical peak performance for each architecture.

4. Bandwidth fraction: ratio of actual data transfer (including read and write) to theoretical bandwidth (on-chip bandwidth).

All these values are relative to respective languages,
i.e. OpenMP for multi-core x86, Cell for QS22, CUDA for Nvidia GPU, and OpenCL for ATI.
However, OpenCL is provided as a reference across different architectures as well.

We calculate the total number of operations in one time step for our FORTRAN version,
including CFL, fluid and magnetic update.
For a single cell in one simulation time step for the box (ignoring $\cal{O}$$({n^2})$),
there are $466$ addition operations,
$598$ subtraction operations,
$1174$ multiplication operations,
$125$ division operations,
and $3$ square root operations.
Since the proportion of division and square root operations are small,
following \cite{nyland2007fast},
we regard their cost as 1 flop each, for simplicity. 
As a result,
the FORTRAN code has 4.62 Giga floating-point operations for a $128^3$ box in each time step, % 4.621276857 
which contains 1 CFL function, 6 fluid update and 6 magnetic update functions. 
Combining the code run times this value one can calculate the actual FLOPS for different architectures.

We calculate the total data load/write for one time step for our FORTRAN version,
including CFL, fluid, magnetic update and memory transpose.
For single-precision in a single cell in one time step,
there are 11 float reads in the CFL function,
10 float reads and 5 float writes in the fluid update,
14 float reads and 6 float writes in the magnetic update,
and 8 float reads and 8 float writes in the memory transpose. 
As a result,
the FORTRAN code has 2.23 GBytes of data transfer (i.e. 1.46 Gbytes read and 0.77 Gbytes write) per time step for a $128^3$ box,
which contains 1 CFL function, 6 fluid and 6 magnetic updates functions, and 4 memory transposes.
Combining the code run times with this value one can calculate the actual bandwidth for different architectures.

Table \ref{table:whole} presents the comparison for different architectures for a box size of $128^3$,
including both the respective program and OpenCL.
Code and fractional speed-up, and FLOPS fraction are included.
We also add the theoretical peak performance for single precision, 
memory bandwidth,
and our practical power consumption in units of watts.
No data for OpenCL on a single core or cell is provided.
The former issue is due to the fact OpenCL treats multi-core x86 as a heterogeneous system and that all the available compute units are used.
The latter is because our OpenCL code still can't run on a Cell cluster,
which may be due to the beta release of OpenCL on Cell.
The power usage for single-core is not available because the Xeon is a multi-core processor.

It can be seen that the CUDA on the Nvidia GPU gets the best speed-up in both code and fractional speed-up.

\subsection{Discussion}

\subsubsection{Speed-up and fractional parameters on different architectures}
The code speed-up quantifies the total gain in performance for different architectures.
The fractional speed-up takes into account the theoretical peak performance comparison
and also the programmer's optimization work relative to the original code.
The FLOPS fraction tells us how many operations are done compared to the peak FLOPS.
The bandwidth fraction tells us what percentage of bandwidth the code occupies.
Comparing the two fractions can give us an idea which one is the bottleneck for the performance.
Our results indicate that CUDA on the Nvidia GPU is a good choice for starting heterogeneous computing.
CUDA on the Nvidia GPU has up to 105 times code speed-up and 2.4 fractional speed-up,
which means that CUDA can hide memory latency well (this partly relates to programmer's optimization work).
CUDA also has the highest FLOPS and bandwidth fraction,
which tells us that CUDA uses its flops computing ability and bandwidth efficiently.

\subsubsection{More detail for CUDA and OpenCL on GPU}
The Nvidia GPU has scalar shader cores and high efficiency of computing.
On the other hand, the ATI GPU has much more cores, which leads to much higher power for floating-point operations,
but with low efficiency of computing.  This may be due to the difficulty
of mapping the algorithm efficiently onto the 4D+1D vector core design.
Since we have both CUDA and OpenCL here, while the latter one can also run on Nvidia GPU,
we did some more comparisons here.

\begin{figure}[!t]
\centering
\includegraphics[width=2.5in]{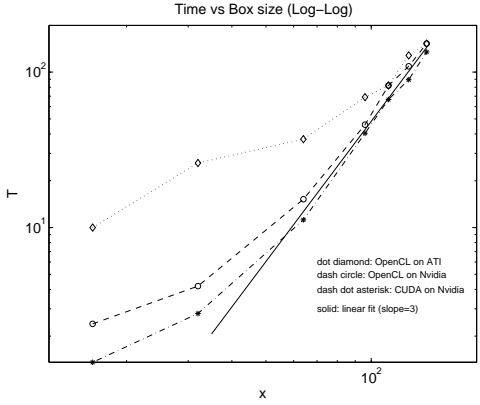}
%\DeclareGraphicsExtensions.
\caption{Time vs box size for GPU comparison.
X axis represents the length of the box;
Y axis represents the time for one time step,
Timings in milli second;
Dot diamond is OpenCL on ATI;
Dash circle is OpenCL on Nvidia;
Dash dot asterisk is CUDA on Nvidia;
Solid is linear fit with slope=3.
}
\label{gpu}
\end{figure}

The comparison for CUDA on Nvidia, OpenCL on Nvidia and OpenCL on ATI is in Fig. \ref{gpu}.
The X axis represents the length for the box, which is in log scale,
and the Y axis represents the time for one time step,
which is in log scale and millisecond time units.
It can be seen that CUDA on Nvidia on smaller box sizes are good.
The OpenCL on ATI catches up to CUDA with increasing box sizes.
The OpenCL on Nvidia performed a little worse than CUDA on Nvidia.
We didn't simulate the box sizes larger than $144^3$ due to the memory limit on ATI.

The fact that the ATI GPU has advantages for larger box sizes is not surprising,
because ATI has double of the shared memory and more spaces for register (128-bit comparing to 32-bit in Nvidia GPU).
The shared memory issue is the bottle neck for magnetic function in CUDA on Nvidia,
and this is also what we saw from our simulation data (not show here).
The bottleneck for the fluid function is register space, but we didn't observe much difference for fluid function,
which may take effect with larger box size ($>160^3$).

\subsubsection{Similar structures for all heterogeneous systems and challenges}
As heterogeneous systems,
both Cell processors and GPU share similar features.
They have a control processor to organize the calculation,
while the computing intensive processors speed up the simulation.
The different roles of the processors require a correspondingly complicated memory system.
 Three levels of memory --- global-shared-local ---
is a typical structure for heterogeneous systems.
Similar to the conventional processors,
there are also memory and bandwidth challenges for heterogeneous systems.

The Cell processor has 32GB main memory, but only 256kB local storage,
which puts a constraint on the size of computing data in the SPEs.
It also limits the range of double/multiple buffering,
which is very important for Cell processors because there are only 8 SPEs. In order to achieve a high performance one has to keep all of them as busy as possible.

The Nvidia GPU has only 1 GB global memory,
and the limited amount of registers and shared memory also puts a constraint on the simulation.
It achieves a bandwidth fraction of 19.1$\%$, which is much higher than the other architectures.  This may hint on targeting future optimization efforts to use bandwidth more efficiently.

The ATI GPU has the same amount of global memory as Nvidia GPUs, but
with twice as much shared memory and the same number of registers (but
128-bit).  However, it has only 20 SIMD units (i.e. workgroup in
OpenCL/block in CUDA) which is less than Nvidia GPU.  As a result, it
may be helpful to send more data into one workgroup than for the
Nvidia GPU.  Still, the limit of three levels memory is a significant
constraint for the simulation.

\subsubsection{Future work}
\begin{itemize}
\item Cell:
Only SIMD and double buffering are included in our simulation,
more can be done to explore the power of Cell/B.E..
\item Nvidia GPU:
The register restriction on fluid update and shared memory restriction on magnetic update limit the occupancy.
Reorganizing the algorithms for them might be helpful to speed up the code.
\item ATI GPU:
The ATI GPU SIMD unit has the problem of low efficiency for vectorized core computing,
We will do more research on this to explore the power of 2.7Tflops ATI GPU.
\item MPI:
We will apply our code to MPI version for use on GPU clusters in the future.
\end{itemize}

\begin{table}
\centering
\caption{
Performance comparison for different architectures; timings in milliseconds.
N-GPU represents GTX 260;
A-GPU represents ATI HD5870;
peak Gflops represents theoretical peak floating-point performance;
peak GB/s represents the theoretical on-chip bandwidth; 
}
%    \advance\extrarowheight-5pt
\begin{tabular}{|c|c|c|c|c|c|}
\hline
Architecture 	& x86(1) & x86(8) & Cell & N-GPU & A-GPU\\
\hline
Respective time & 8770 	& 1315 	& 864 	& 83 	& 128 	\\
OpenCL time 	& N/A 	& 6435 	& N/A  	& 109 	& 128 	\\
Peak Gflops 	& 17 	& 136 	& 409.6 & 748.8 & 2720 	\\
Peak GB/s 	& 19.2 	& 19.2 	& 204.8 & 141 	& 153.6 \\
Power(Watts) 	& N/A 	& 170 	& 440 	& 370 	& 360 	\\
Code speed-up 	& 1.0 	& 6.7 	& 10.2 	& 105.7	& 68.5 	\\
Fractional speed-up & 1.0 & 0.83 & 0.42	& 2.40 	& 0.43	\\
FLOPS fraction 	& 3.1\% & 2.6\% & 1.3\% & 7.4\%  & 1.3\% \\	% 4.62 GFLOP
Bandwidth fraction & 1.3\% & 8.8\% & 1.3\% & 19.1\% & 11.3\% \\	% 2.23 GB
% 27 w idle for ati
\hline
\end{tabular}
\label{table:whole}
\end{table}

\begin{figure*}[!t]
\centering
\includegraphics[width=6in]{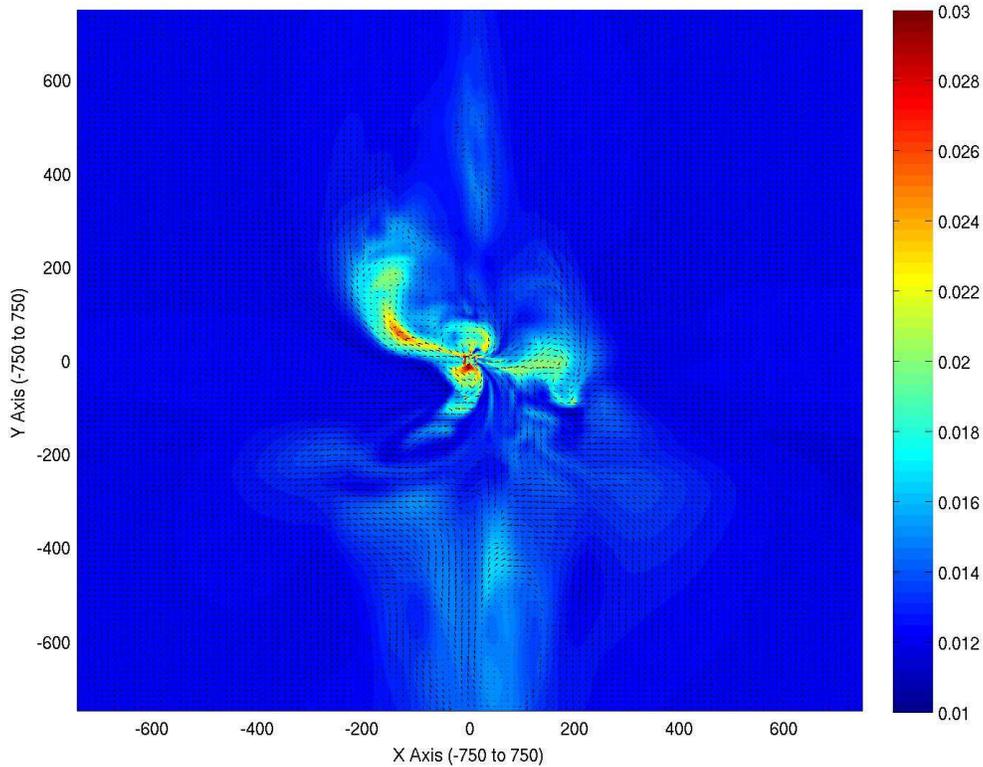}
%\DeclareGraphicsExtensions.
\caption{2D snapshot of black hole accretion simulation.
Color represents entropy and arrows represent magnetic field}
\label{blackhole}
\end{figure*}

To conclude this discussion, 
we present a simulation 2D snapshot (Fig. \ref{blackhole}) from our black hole accretion simulation.

\section{Conclusion}

We presented magneto-hydrodynamics simulations on heterogeneous
systems, e.g. Cell/B.E., Nvidia and ATI GPU.  These heterogeneous
systems share a similar structure that they all have a control
processor for mission management and many computing intensive
processors for calculations.  Correspondingly, the memory system is
also complicated, which is a challenge for programmers.  We present
the results on different architectures for comparison; 10 times, 105 times
and 68 times speed-up for Cell, Nvidia, and ATI GPU were achieved.  The
CUDA on Nvidia GPU has the best performance on both code and fractional
speed-up, and the ATI GPU improves with larger size simulation.
The 2.4 fractional speed-up for CUDA on Nvidia GPU shows that a greater percentage of peak theoretical performance compared to x86 architecture was achieved.

These performance numbers were obtained with an algorithm which was
directly translated from a CPU code.  Designing algorithms with
heterogeneous architectures in mind may also improve performance.


\section*{Acknowledgement}

We would like to thank Jonathan Dursi, Wenda Han, Qi Liu, Harald Pfeiffer, Scott
Rostup, Daniele P. Scarpazza for helpful suggestions.  We thank Scott
Rostup, Hsi-Yu Schive, Tomoyoshi Shimobaba for providing their code
for our reference, and Hon-Cheng Wong for providing the detail of
their simulation results, and Kiyoshi Wesley Masui for reviewing the
draft, and Gojko Vujanovic for providing the power consumption
measurement.  We acknowledge IBM TJ Watson Research Center for
providing the IBM QS22 Cell blade.  Part of Cell computations were
performed on the Cell cluster at the SciNet HPC Consortium. SciNet is
funded by: the Canada Foundation for Innovation under the auspices of
Compute Canada; the Government of Ontario; Ontario Research Fund -
Research Excellence; and the University of Toronto.

The work of BP is supported by the MITACS ACCELERATE Scholarship.



\bibliographystyle{IEEEtran}
\bibliography{IEEEfull,bpang}
%\begin{thebibliography}{1}


%\end{thebibliography}

\end{document}